\documentclass[a4paper]{article}

\usepackage[english]{babel}
\usepackage[utf8x]{inputenc}
\usepackage{amsmath}
\usepackage[pdftex]{graphicx}
\usepackage[colorinlistoftodos]{todonotes}

\usepackage[caption=false,font=footnotesize]{subfig}
\usepackage{booktabs}
\usepackage{multirow}
\usepackage{latexsym, amssymb, verbatim}
\usepackage{ulem}
\usepackage{color}
\usepackage{indentfirst}
\newtheorem{definition}{Definition}
\newtheorem{theorem}{Theorem}
\allowdisplaybreaks

\title{A Novel Admission Control Model in Cloud Computing}
\author{Yunlong He$^1$, Jun Huang$^1$, Qiang Duan$^2$, Zi Xiong$^1$,\\
       Juan Lv$^1$, Yanbing Liu$^3$\\
       $^1$School of Communication and Information Engineering,\\
        Chongqing University of Posts and Telecommunications,\\
        Chongqing, China 400065\\
       Email: xiaoniuadmin@gmail.com\\
       $^2$Info. Science and Technology Department, The\\ Pennsylvania
        State University, Abington, Pennsylvania 19001\\
       Email: qduan@psu.edu\\
       $^3$School of Computer Science and Technology,\\
        Chongqing University of Posts and Telecommunications,\\
        Chongqing, China 400065\\
       Email: liuyb@cqupt.edu.cn
}

\begin{document}
\maketitle

\begin{abstract}

With the rapid development of Cloud computing technologies and wide adopt of Cloud services and applications, QoS provisioning in Clouds becomes an important research topic. In this paper, we propose an admission control mechanism for Cloud computing. In particular we consider the high volume of simultaneous requests for Cloud services and develop admission control for aggregated traffic flows to address this challenge. By employ network calculus, we determine effective bandwidth for aggregate flow, which is used for making admission control decision. In order to improve network resource allocation while achieving Cloud service QoS, we investigate the relationship between effective bandwidth and equivalent capacity. We have also conducted extensive experiments to evaluate performance of the proposed admission control mechanism.

\end{abstract}

\section{Introduction}

Recently the emerging Cloud computing has been developing very quickly \cite{mishra2013, armbrust2010, ashraf2013}. With the rapid development of Cloud computing technologies and wide adoption of Cloud-based applications, the huge amount of traffic generated by a large number of users for accessing Cloud services bring in a series challenge to the Internet. The best-effort service model in the current Internet cannot meet users’ requirements for Quality of Service (QoS). Call Admission Control (CAC) offers an effective approach to controlling network traffic and avoiding network congestion; thus facilitating QoS provisioning for Cloud services. A key component of CAC is to determine the minimal amount of resources required for meeting application performance requirement. Network calculus offers an effective method for such worst-case analysis.

In this paper we will first give a brief overview of related work on Cloud admission control in Section 2 and provide an introduction to network calculus in Section 3. A model of admission control for Cloud services is proposed in Section 4. Then in Section 5 we develop a technique for admission control of aggregate flow and examine the relation between effective bandwidth and equivalent capacity. Experimental results are reported in Section 6 for evaluating performance of the proposed admission control technology.

\section{Related Work}

Network calculus was initially invented by Chang \cite{chang2000} and Cruz \cite{cruz1, cruz2} and then further developed by other researchers (e.g., \cite{parekh1993, agrawal1999, jiang2008, le1998}) into an effective quantitative tool for analyzing network performance. Network calculus uses arrival curve and service curve to determine some key QoS factors of networking systems such as delay and backlog \cite{ciucu2012, bouillard2008}. Compared to traditional queuing analysis methods, network calculus can provide performance bounds for networking systems to obtain work-case performance, which allows strict QoS guarantee \cite{bisti2012, bouillard2010}. Network calculus has been widely applied in network performance evaluation, through which tight performance bounds can be obtained for making admission decisions \cite{shan1998, schmitt2008}.

Cloud admission control has started attracting more attention of the research community \cite{ashraf2013, wu2012}. Ashraf et al. \cite{ashraf2013} and his colleagues proposed a single flow-based admission control method for Cloud services. However, wide the rapid development of Cloud computing a large number of users may send service requests in parallel. Therefore, single flow-based admission control limits the scalability of Cloud service provisioning. Le Boudec et al. \cite{le2001} proposed the concepts of delay-based Effective Bandwidth (EB) and backlog-based Equivalent Capacity (EC), which can be used for network call admission control. However, application of EB and EC in Cloud admission control is still an open issue.

In this paper we propose a network and system model of admission control for Cloud services. In order to address the challenge brought in by the large number of parallel cloud service requests, we particularly study admission control for aggregate flow. We develop a technique for determining the effective bandwidth for aggregate flow for making admission decisions. In order to improve resource allocation as well as providing QoS guarantee, we also examine the relationship between effective bandwidth and equivalent bandwidth, and especially analyze such relationship for aggregate flow. We conduct extensive numerical experiments to study features of effective bandwidth of various flows and evaluate the proposed admission control scheme under various delay requirements.

\section{Network Calculus Theory}

We now give the definition and theorem of some of the notions in network calculus that will be needed in the rest of the paper. The detailed descriptions of these concepts can be found in \cite{le2001}.

\begin{definition}
(Arrival Curve). Given a general increasing function $\alpha$, we say that a flow $R$ has $\alpha$ as an arrival curve, or R is $\alpha$-smooth, if and only if $R$ meets one of the following two equivalent conditions for $\forall t \ge 0,s \le t$:

\begin{eqnarray}
  &&R(t) - R(s) \le \alpha (t - s)\\
  &&R(t) \le (R \otimes \alpha )(t)
\end{eqnarray}

\end{definition}

\noindent where $\otimes$ is the min-plus convolution and is given as follows:

\begin{equation}\label{}
  (R \otimes \alpha )(t) = \left\{ {\begin{array}{*{20}{c}}
  {{{\inf }_{0 \le s \le t}}\{ R(s) + \alpha (t - s)\} ,{\kern 1pt} {\kern 1pt} {\kern 1pt} {\kern 1pt} {\kern 1pt} {\kern 1pt} {\kern 1pt} {\kern 1pt} {\kern 1pt} {\kern 1pt} {\kern 1pt} t \ge 0}\\
  {0,{\kern 1pt} {\kern 1pt} {\kern 1pt} {\kern 1pt} {\kern 1pt} {\kern 1pt} {\kern 1pt} {\kern 1pt} {\kern 1pt} {\kern 1pt} {\kern 1pt} t < 0}
  \end{array}} \right.
\end{equation}

\begin{definition}
(Service Curve). Consider a system S and a flow through S with input and output function R and ${R^*}$. We say that S offers to the flow a service curve $\beta$ if and only if $\beta$ is wide sense increasing, and $\beta (0) = 0$, for any time instant t satisfies that:

\begin{equation}\label{}
  {R^*}(t) \ge (R \otimes \beta )(t)
\end{equation}

\end{definition}

\begin{definition}
(Effective Bandwidth). For a flow with an arrival curve $\alpha$, the effective bandwidth ${e_D}(\alpha )$ of the flow is defined to be the bit rate required to serve the flow in a work conserving manner, with a delay constraint D. That is,
\begin{equation}\label{}
  {e_D}(\alpha ) = {\sup _{s \ge 0}}\left\{ {\frac{{\alpha (s)}}{{s + D}}} \right\}
\end{equation}

\end{definition}

Regarding the effective bandwidths of aggregate flow, we may assume ${D_i} = D$ for any $i$ (where $i$ represents different types of flow), then we have:

\begin{equation}\label{eq:3.6}
  {e_D}(\sum\nolimits_i {{\alpha _i}} ) \le \sum\nolimits_i {{e_D}} ({\alpha _i})
\end{equation}

If this assumption causes trouble, then we may provide a guaranteed delay for every cloud service flow by letting $D = \min \{ {D_i}\}$.

\begin{definition}
(Equivalent Capacity). Considering a node S with buffer size B and a flow going through S with arrival curve $\alpha$, the equivalent capacity, ${f_B}(\alpha )$, is defined as follows:

\begin{equation}\label{}
  {f_B}(\alpha ) = {\sup _{s > 0}}\left\{ {\frac{{\alpha (s) - B}}{s}} \right\}
\end{equation}
\end{definition}

Note that the equivalent capacity of aggregate flow has the following property:

\begin{equation}\label{eq:3.8}
  {f_B}(\alpha ) \le \sum\nolimits_i {{f_{{B_i}}}} ({\alpha _i})
\end{equation}

\noindent where $\alpha {\rm{ = }}\sum\nolimits_i {{\alpha _i}} ,B = \sum\nolimits_i {{B_i}}$.

\begin{theorem}
(Backlog Bound). Assume a flow, constrained by arrival curve $\alpha$, traverses a system that offers a service curve $\beta$. The backlog $R(t) - {R^*}(t)$ for all t satisfies:

\begin{equation}\label{}
  \begin{array}{l}
    R(t) - {R^*}(t) \le v(\alpha ,\beta )\\
    {\kern 1pt} {\kern 1pt} {\kern 1pt} {\kern 1pt} {\kern 1pt} {\kern 1pt} {\kern 1pt} {\kern 1pt} {\kern 1pt} {\kern 1pt} {\kern 1pt} {\kern 1pt} {\kern 1pt} {\kern 1pt} {\kern 1pt} {\kern 1pt} {\kern 1pt} {\kern 1pt} {\kern 1pt} {\kern 1pt} {\kern 1pt} {\kern 1pt} {\kern 1pt} {\kern 1pt} {\kern 1pt} {\kern 1pt} {\kern 1pt} {\kern 1pt} {\kern 1pt} {\kern 1pt} {\kern 1pt} {\kern 1pt} {\kern 1pt} {\kern 1pt} {\kern 1pt} {\kern 1pt} {\kern 1pt} {\kern 1pt} {\kern 1pt} {\kern 1pt} {\kern 1pt} {\kern 1pt} {\kern 1pt} {\kern 1pt} {\kern 1pt} {\kern 1pt} {\kern 1pt} {\kern 1pt} {\kern 1pt} {\kern 1pt}  = {\sup _{s \ge 0}}\left\{ {\alpha (s) - \beta (s)} \right\}\\
    {\kern 1pt} {\kern 1pt} {\kern 1pt} {\kern 1pt} {\kern 1pt} {\kern 1pt} {\kern 1pt} {\kern 1pt} {\kern 1pt} {\kern 1pt} {\kern 1pt} {\kern 1pt} {\kern 1pt} {\kern 1pt} {\kern 1pt} {\kern 1pt} {\kern 1pt} {\kern 1pt} {\kern 1pt} {\kern 1pt} {\kern 1pt} {\kern 1pt} {\kern 1pt} {\kern 1pt} {\kern 1pt} {\kern 1pt} {\kern 1pt} {\kern 1pt} {\kern 1pt} {\kern 1pt} {\kern 1pt} {\kern 1pt} {\kern 1pt} {\kern 1pt} {\kern 1pt} {\kern 1pt} {\kern 1pt} {\kern 1pt} {\kern 1pt} {\kern 1pt} {\kern 1pt} {\kern 1pt} {\kern 1pt} {\kern 1pt} {\kern 1pt} {\kern 1pt} {\kern 1pt} {\kern 1pt} {\kern 1pt} {\kern 1pt}  = (\alpha  \emptyset \beta )(0)
  \end{array}
\end{equation}

\end{theorem}

\noindent where $v(\alpha ,\beta )$ is $vertical$ $deviation$ between $\alpha$ and $\beta$, the $\emptyset$ is the min-plus deconvolution and is given as follows:

\begin{equation}\label{}
  (\alpha  \emptyset \beta )(t) = {\sup _{v \ge 0}}\left\{ {\alpha (t + v) - \beta (v)} \right\}
\end{equation}

\begin{theorem}
(Delay Bound). Assume a flow, constrained by arrival curve $\alpha$, traverses a system that offers a service curve of $\beta$. The  delay d(t) for all t satisfies:

\begin{equation}\label{}
  \begin{array}{l}
    d(t) \le h(\alpha ,\beta )\\
    {\kern 1pt} {\kern 1pt} {\kern 1pt} {\kern 1pt} {\kern 1pt} {\kern 1pt} {\kern 1pt} {\kern 1pt} {\kern 1pt} {\kern 1pt} {\kern 1pt} {\kern 1pt} {\kern 1pt} {\kern 1pt} {\kern 1pt} {\kern 1pt} {\kern 1pt} {\kern 1pt}  = {\sup _{s \ge 0}}\left\{ {\inf \left\{ {T \ge 0:\alpha (s) \le \beta (s + T)} \right\}} \right\}
  \end{array}
\end{equation}

\end{theorem}

\noindent where $h(\alpha ,\beta )$ is the $horizontal$ $deviation$ between $\alpha$ and $\beta$.

\section{Admission Control Model for Cloud Services}

\begin{figure}[h]
\centering
\subfloat[\texttt{Network Model}]{\includegraphics[width=4in]{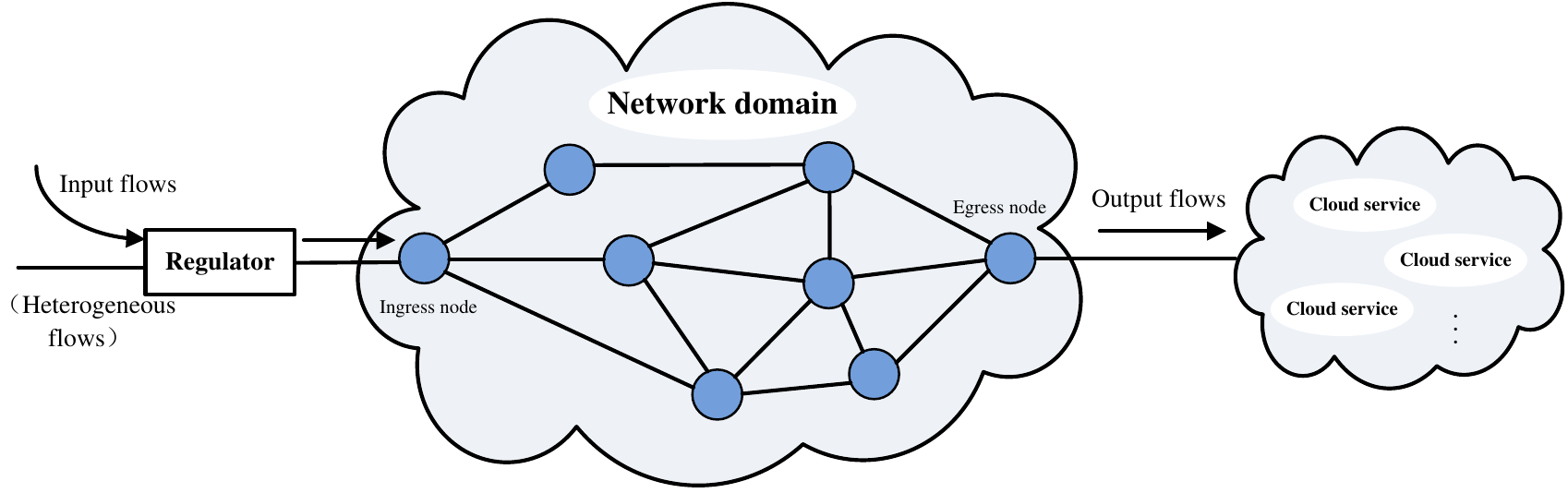}}\\
\subfloat[\texttt{System Model}]{\includegraphics[width=4in]{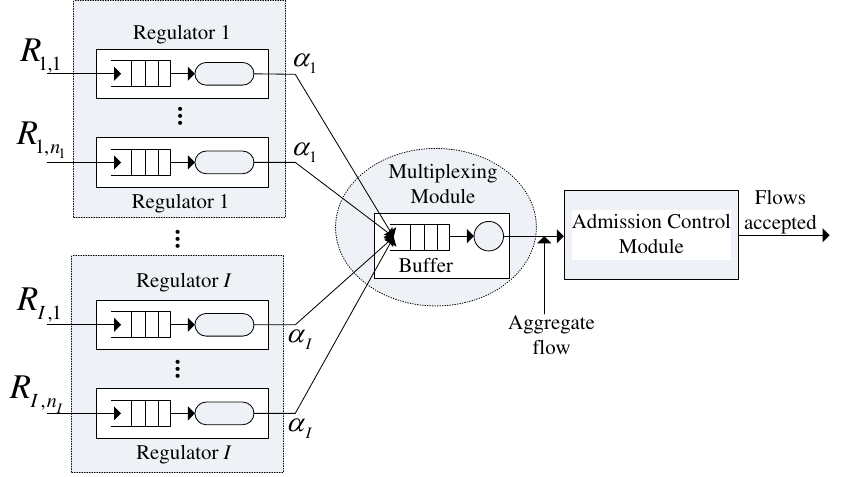}}\\
\caption{Admission Control Model for Cloud Services}
\label{fig1}
\end{figure}

The admission control model we proposed for Cloud services is shown in Figure \ref{fig1} which consists of a network model and a system model.

Figure \ref{fig1}(a). is the network topology by which users request the cloud services. The process of users’ requesting and accepting cloud services must be done via the traditional Internet. Figure \ref{fig1}(b) is the system model for admission control. The heterogeneous flows of user requests for cloud services are first shaped by the regulator; then the shaped flows are multiplexed through the FIFO multiplexing module; and finally the output from the multiplexing module, which is the aggregate flow, goes through the phase of admission control producing the accepted flow. The multiplexing module and admission control module are deployed at the position of the ingress node in Figure \ref{fig1}(a), so that those flows that are accepted will enter the network domain.

As shown in Figure \ref{fig1}(b), the heterogeneous flows share a common buffer when they are multiplexed, which indicates that this mechanism requires less bandwidth resources than that where each flow is allocated a fixed size buffer. This is exactly the situation described by the inequalities in (\ref{eq:3.6}) and (\ref{eq:3.8}). In other words, the admission control with respect to aggregate flow can admit more cloud services than that of per flow when the bandwidth of the ingress node is a constant.

Heterogeneous flows’ bursts are smoothed through the regulator before being multiplexed. We consider the case where the output flows of regulators are constrained by the traffic specification $T$-$SPEC (p, M, r, b)$. $T$-$SPEC (p, M, r, b)$ is the shaping curve of the regulator, and is the arrival curve of the output flows of regulators as well, where parameters $p$, $M$, $r$, $b$ are peak rate, maximum packet size, sustainable rate (average rate), and burst tolerance of a flow, respectively. The specific constraint function is as follows:

\begin{equation}\label{eq:4.12}
  \alpha (t) = \left\{ {\begin{array}{*{20}{c}}
  {\min (pt + M,rt + b),{\kern 1pt} {\kern 1pt} {\kern 1pt} {\kern 1pt} {\kern 1pt} {\kern 1pt} {\kern 1pt} {\kern 1pt} {\kern 1pt} {\kern 1pt} {\kern 1pt} {\kern 1pt} {\kern 1pt} {\kern 1pt} t \ge 0}\\
  {0{\kern 1pt} ,{\kern 1pt} {\kern 1pt} {\kern 1pt} {\kern 1pt} {\kern 1pt} {\kern 1pt} {\kern 1pt} {\kern 1pt} {\kern 1pt} {\kern 1pt} {\kern 1pt} {\kern 1pt} {\kern 1pt} {\kern 1pt} {\kern 1pt} t < 0}
\end{array}} \right.
\end{equation}

\section{Admission Control for Cloud Services}

In this section, we will introduce the proposed admission control approach for cloud services in details.

\subsection{Relationship between Effective Bandwidth and Equivalent Capacity}

Effective bandwidth and equivalent capacity are two critical concepts, which are defined from the perspectives of cloud services and networking, respectively. However, they are not independent of each other. As a matter of fact, the close relationship between them is the foundation to optimize the network performance. From Theorem 1 and Theorem 2, we are ready to derive their relationship expressed in Theory 3 as follows.

\begin{theorem}
A flow with the arrival curve $\alpha$, goes through node $S$ with the buffer size $B$. 1)  given the delay constraint $D$, this flow's effective bandwidth is equal to the equivalent capacity as long as the buffer size $B$ meets $B=h(D)$; 2) given the buffer size $B$, this flow’s effective bandwidth is equal to the equivalent capacity as long as the flow’s delay constraint $D$ meets $D=g(B)$, where $h(D) = \sup_{s \ge 0}\{\alpha(s) - e_D(\alpha) s\}$, $g(B)=\sup_{s \ge 0}\{\inf\{ T \ge 0: \alpha(s) \le f_B(\alpha) (s+T)\}\}$.
\end{theorem}

In essence, both EB and EC represent the service rates of a node given a specific constraint. This implies that with a specific constraint, a
node is able to process the same amount of cloud services if the effective bandwidth of flow is equal to the equivalent capacity of the flow, otherwise, the amount will be determined by the minimum value  between effective bandwidth and equivalent capacity.

Theorem 3 can be mathematically approved. 1) Given the delay constraint $D$, The node $S$ provides the service rate for a flow is ${e_D}(\alpha )$. Theorem 3 can be expressed as $e_D(\alpha) \mathop  = \limits^{B = h(D)} f_B(\alpha)$ where ${e_D}(\alpha )$ and ${f_B}(\alpha )$ represent the effective bandwidth and equivalent capacity, respectively. Given $B = h(D)$, the flow’s equivalent capacity is:

\begin{equation}
{f_B}(\alpha ) = {\sup _{s \ge 0}}\{ \frac{{\alpha (s) - B}}{s}\}  = {\sup _{s \ge 0}}\{ \frac{{\alpha (s) - {{\sup }_{t \ge 0}}\{ \alpha (t) - {e_D}(\alpha )t\} }}{s}\}
\label{eq:5.1}
\end{equation}

Since

\begin{equation}
  {\sup _{t \ge 0}}\{ \alpha (t) - {e_D}(\alpha )t\}  \ge \alpha (t) - {e_D}(\alpha )t
\label{eq:5.2}
\end{equation}

From (\ref{eq:5.2}),

\begin{equation}
  \frac{{\alpha (s) - {{\sup }_{t \ge 0}}\{ \alpha (t) - {e_D}(\alpha )t\} }}{s} \le {e_D}(\alpha )
\label{eq:5.3}
\end{equation}

As we can see, we simply perform scaling operations in the domain of the function. This indicates that the maximum value of, i.e., the upper bound of $\frac{{\alpha (s) - {{\sup }_{t \ge 0}}\{ \alpha (t) - {e_D}(\alpha )t\} }}{s}$ is ${e_D}(\alpha )$. From (\ref{eq:5.3}), we conclude that ${f_B}(\alpha ) = {e_D}(\alpha )$.

2)  Given the buffer size $B$, the node $S$ provides the service rate for a flow is ${f_B}(\alpha )$. With this scenario, Theory 3 can be expressed as ${f_B}(\alpha )\mathop  = \limits^{D = g(B)} {e_D}(\alpha )$, where ${e_D}(\alpha )$ and ${f_B}(\alpha )$ represent the effective bandwidth and equivalent capacity, respectively. Given $D = g(B)$, the flow’s effective bandwidth is :

\begin{equation}
{e_D}(\alpha ) = {\sup _{s \ge 0}}\{ \frac{{\alpha (s)}}{{s + D}}\}  = {\sup _{s \ge 0}}\{ \frac{{\alpha (s)}}{{s + {{\sup }_{t \ge 0}}\{ \inf \{ T \ge 0:\alpha (t) \le {f_B}(\alpha )(t + T)\} \} }}\}
\label{eq:5.4}
\end{equation}

Since,
\begin{equation}
\begin{array}{l}
{\sup _{t \ge 0}}\{ \inf \{ T \ge 0:\alpha (t) \le {f_B}(\alpha )(t + T)\} \} \\
 = {\sup _{t \ge 0}}\{ \inf \{ T \ge 0:T \ge \frac{{\alpha (t) - {f_B}(\alpha )t}}{{{f_B}(\alpha )}}\} \} \\
 = {\sup _{t \ge 0}}\{ \frac{{\alpha (t) - {f_B}(\alpha )t}}{{{f_B}(\alpha )}}\} \\
 \ge \frac{{\alpha (t) - {f_B}(\alpha )t}}{{{f_B}(\alpha )}}
\label{eq:5.5}
\end{array}
\end{equation}

From (\ref{eq:5.5}), we derive:

\begin{equation}
\frac{{\alpha (s)}}{{s + {{\sup }_{t \ge 0}}\{ \inf \{ T \ge 0:\alpha (t) \le {f_B}(\alpha )(t + T)\} \} }} \le \frac{{\alpha (s)}}{{s + \frac{{\alpha (s) - {f_B}(\alpha )s}}{{{f_B}(\alpha )}}}} = {f_B}(\alpha )
\label{eq:5.6}
\end{equation}

Therefore, from (\ref{eq:5.6}), we conclude that ${e_D}(\alpha ) = {f_B}(\alpha )$. This completes the proof.

\subsection{Characteristics of Aggregate Flow}

After examining the EB and EC of a single flow, now we turn to the characteristics of an aggregate flow. Specifically, we will analyze the calculation of effective bandwidth of the aggregate flow, and the required buffer size $B$ to satisfy the specified delay constraint while serving the same amount of cloud services.

\subsubsection{Effective Bandwidth of Aggregate Flow}

Suppose there are $I$ types of cloud services, and the multiplexing module multiplexes $n_i$ flows of type $i$, where $n_i$ is the number of cloud services of type $i(i=1,2,…,I)$. Every flow has $T$-${SPEC(p_i, M_i, r_i, b_i)}$ as an arrival curve. The multiplexing process is as shown in Figure 1(b), and a fixed, but arbitrary, delay constraint $D$ is set for each flow. The Effective Bandwidth for aggregate flow is given by the following formula:

\[{e_D}(\sum\limits_{i = 1}^I {{n_i}{\alpha _i}} )\]

where ${\alpha_i=T}$-${SPEC(p_i, M_i, r_i, b_i)}$ and $\sum\limits_{i = 1}^I {{n_i}{\alpha _i}} $ is the arrival curve of the aggregate flow.

\begin{figure}[h]
    \centering
    \includegraphics[width=3in]{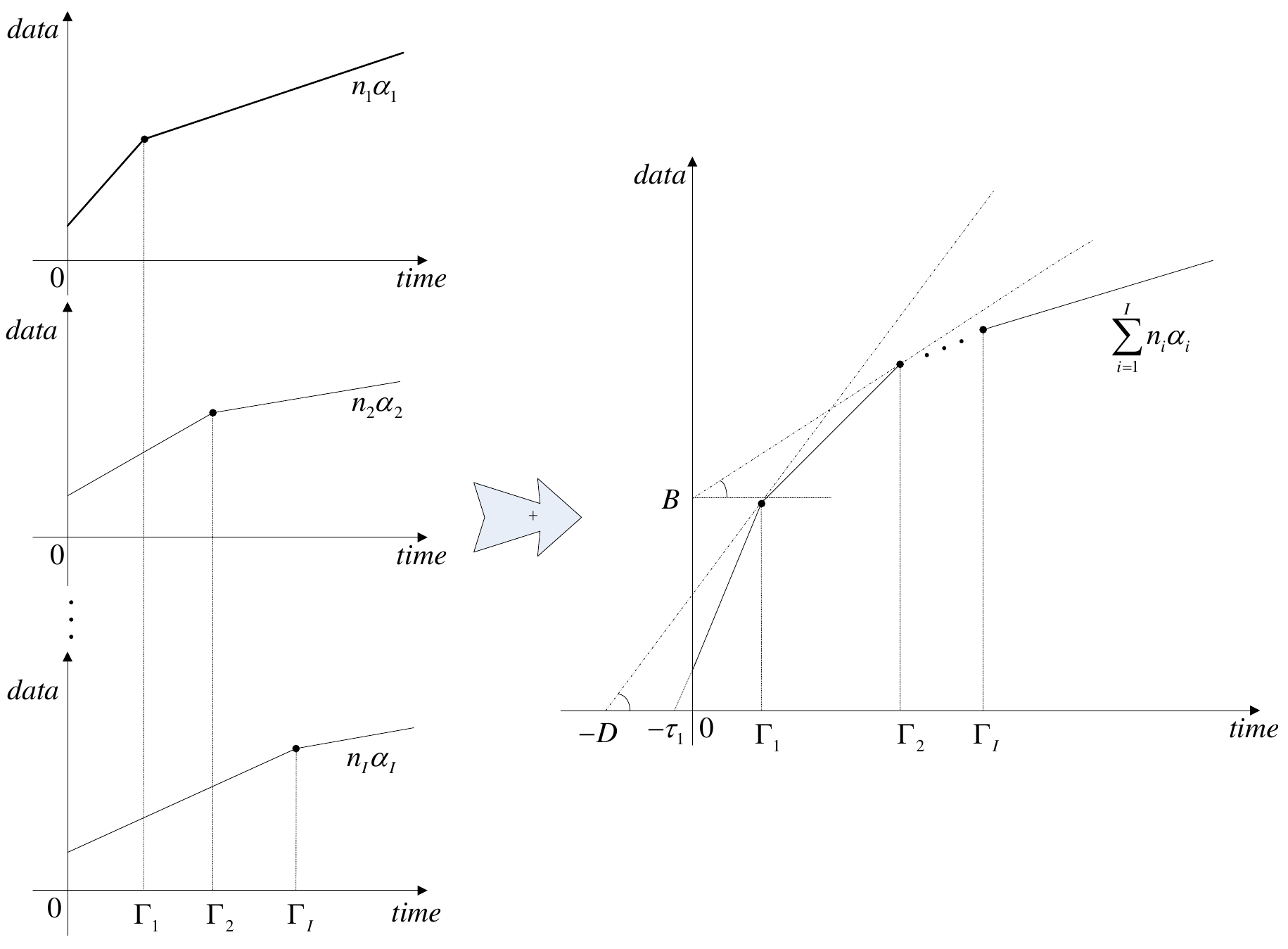}
    \caption{Calculation of EB for Aggregate Flow}
    \label{fig2}
\end{figure}

According to the definition in (5), we design an approach to calculating the effective bandwidth of an aggregate flow, as illustrated in Figure \ref{fig2}. Let ${\Gamma _i} = \frac{{{b_i} - {M_i}}}{{{p_i} - {r_i}}}$ and assume ${\Gamma _1} \le {\Gamma _2} \le  \cdots  \le {\Gamma _I}$.

As we know, the EB of an aggregate flow is the slope of the corresponding arrival curve at horizontal axis $time=-D$, while the EC is the slope of the arrival curve at vertical axis $data=B$.

Therefore, the EB of an aggregate flow is represented as:

\begin{equation}
\begin{array}{l}
{e_D}(\sum\limits_{i = 1}^I {{n_i}{\alpha _i}} ) = \max \{ {e_1},{e_2},{e_3}, \cdots ,{e_{I + 1}},{e_{I + 2}}\} \\
{\kern 1pt} {\kern 1pt} {\kern 1pt} {\kern 1pt} {\kern 1pt} {\kern 1pt} {\kern 1pt} {\kern 1pt} {\kern 1pt} {\kern 1pt} {\kern 1pt} {\kern 1pt} {\kern 1pt} {\kern 1pt} {\kern 1pt} {\kern 1pt} {\kern 1pt} {\kern 1pt} {\kern 1pt} {\kern 1pt} {\kern 1pt} {\kern 1pt} {\kern 1pt} {\kern 1pt} {\kern 1pt} {\kern 1pt} {\kern 1pt} {\kern 1pt} {\kern 1pt} {\kern 1pt} {\kern 1pt} {\kern 1pt} {\kern 1pt} {\kern 1pt} {\kern 1pt} {\kern 1pt} {\kern 1pt} {\kern 1pt} {\kern 1pt} {\kern 1pt} {\kern 1pt} {\kern 1pt} {\kern 1pt} {\kern 1pt} {\kern 1pt} {\kern 1pt}  = \left\{ {\begin{array}{*{20}{c}}
{{e_1}{\kern 1pt} ,{\kern 1pt} {\kern 1pt} {\kern 1pt} {\kern 1pt} {\kern 1pt} {\kern 1pt} {\kern 1pt} {\kern 1pt} {\kern 1pt} {\kern 1pt} {\kern 1pt} {\kern 1pt} {\kern 1pt} {\kern 1pt} {\kern 1pt} {\kern 1pt} {\kern 1pt} {\kern 1pt} {\kern 1pt} {\kern 1pt} 0 < D \le {\tau _1}}\\
{{e_2},{\kern 1pt} {\kern 1pt} {\kern 1pt} {\kern 1pt} {\kern 1pt} {\kern 1pt} {\kern 1pt} {\kern 1pt} {\kern 1pt} {\kern 1pt} {\kern 1pt} {\kern 1pt} {\kern 1pt} {\kern 1pt} {\kern 1pt} {\kern 1pt} {\tau _1} < D \le {\tau _2}}\\
{{e_3},{\kern 1pt} {\kern 1pt} {\kern 1pt} {\kern 1pt} {\kern 1pt} {\kern 1pt} {\kern 1pt} {\kern 1pt} {\kern 1pt} {\kern 1pt} {\kern 1pt} {\kern 1pt} {\kern 1pt} {\kern 1pt} {\kern 1pt} {\kern 1pt} {\kern 1pt} {\tau _2} < D \le {\tau _3}}\\
 \vdots \\
{{e_{I + 1}}{\kern 1pt} ,{\kern 1pt} {\kern 1pt} {\kern 1pt} {\kern 1pt} {\kern 1pt} {\kern 1pt} {\kern 1pt} {\kern 1pt} {\kern 1pt} {\tau _I} < D \le {\tau _{I + 1}}}\\
{{e_{I + 2}},{\kern 1pt} {\kern 1pt} {\kern 1pt} {\kern 1pt} {\kern 1pt} {\kern 1pt} {\kern 1pt} {\kern 1pt} {\kern 1pt} {\kern 1pt} {\kern 1pt} {\kern 1pt} {\kern 1pt} {\kern 1pt} {\kern 1pt} {\kern 1pt} {\kern 1pt} {\kern 1pt} {\kern 1pt} {\kern 1pt} {\kern 1pt} {\kern 1pt} {\kern 1pt} {\kern 1pt} D > {\tau _{I + 1}}}
\end{array}} \right.
\end{array}
\label{eq:2.7}
\end{equation}

where,

\[\begin{array}{l}
{e_1} = {{\sum\limits_{i = 1}^I {{n_i}{M_i}} } \mathord{\left/
 {\vphantom {{\sum\limits_{i = 1}^I {{n_i}{M_i}} } D}} \right.
 \kern-\nulldelimiterspace} D},\\
{e_2} = {{[(\sum\limits_{i = 1}^I {{n_i}{p_i}} ){\Gamma _1} + \sum\limits_{i = 1}^I {{n_i}{M_i}} ]} \mathord{\left/
 {\vphantom {{[(\sum\limits_{i = 1}^I {{n_i}{p_i}} ){\Gamma _1} + \sum\limits_{i = 1}^I {{n_i}{M_i}} ]} {({\Gamma _1} + D)}}} \right.
 \kern-\nulldelimiterspace} {({\Gamma _1} + D)}},\\
{e_3} = {{[({n_1}{r_1} + \sum\limits_{i = 2}^I {{n_i}{p_i}} ){\Gamma _2} + {n_1}{b_1} + \sum\limits_{i = 2}^I {{n_i}{M_i}} ]} \mathord{\left/
 {\vphantom {{[({n_1}{r_1} + \sum\limits_{i = 2}^I {{n_i}{p_i}} ){\Gamma _2} + {n_1}{b_1} + \sum\limits_{i = 2}^I {{n_i}{M_i}} ]} {({\Gamma _2} + D)}}} \right.
 \kern-\nulldelimiterspace} {({\Gamma _2} + D)}},\\
{e_{I + 1}} = {{[(\sum\limits_{i = 1}^{I - 1} {{n_i}{r_i}}  + {n_I}{p_I}){\Gamma _I} + \sum\limits_{i = 1}^{I - 1} {{n_i}{b_i}}  + {n_I}{M_I}]} \mathord{\left/
 {\vphantom {{[(\sum\limits_{i = 1}^{I - 1} {{n_i}{r_i}}  + {n_I}{p_I}){\Gamma _I} + \sum\limits_{i = 1}^{I - 1} {{n_i}{b_i}}  + {n_I}{M_I}]} {({\Gamma _I} + D)}}} \right.
 \kern-\nulldelimiterspace} {({\Gamma _I} + D)}},\\
{e_{I + 2}} = \sum\limits_{i = 1}^I {{n_i}{r_i}} ,\\
{\tau _1} = {{\sum\limits_{i = 1}^I {{n_i}{M_i}} } \mathord{\left/
 {\vphantom {{\sum\limits_{i = 1}^I {{n_i}{M_i}} } {\sum\limits_{i = 1}^I {{n_i}{p_i}} }}} \right.
 \kern-\nulldelimiterspace} {\sum\limits_{i = 1}^I {{n_i}{p_i}} }}{\kern 1pt} ,\\
{\tau _2} = {{({n_1}{b_1} + \sum\limits_{i = 2}^I {{n_i}{M_i}} )} \mathord{\left/
 {\vphantom {{({n_1}{b_1} + \sum\limits_{i = 2}^I {{n_i}{M_i}} )} {({n_1}{r_1} + \sum\limits_{i = 2}^I {{n_i}{p_i}} )}}} \right.
 \kern-\nulldelimiterspace} {({n_1}{r_1} + \sum\limits_{i = 2}^I {{n_i}{p_i}} )}},\\
{\tau _3} = {{(\sum\limits_{i = 1}^2 {{n_i}{b_i}}  + \sum\limits_{i = 3}^I {{n_i}{M_i}} )} \mathord{\left/
 {\vphantom {{(\sum\limits_{i = 1}^2 {{n_i}{b_i}}  + \sum\limits_{i = 3}^I {{n_i}{M_i}} )} {(\sum\limits_{i = 1}^2 {{n_i}{r_i}}  + \sum\limits_{i = 3}^I {{n_i}{p_i}} )}}} \right.
 \kern-\nulldelimiterspace} {(\sum\limits_{i = 1}^2 {{n_i}{r_i}}  + \sum\limits_{i = 3}^I {{n_i}{p_i}} )}},\\
{\tau _I} = {{(\sum\limits_{i = 1}^{I - 1} {{n_i}{b_i}}  + {n_I}{M_I})} \mathord{\left/
 {\vphantom {{(\sum\limits_{i = 1}^{I - 1} {{n_i}{b_i}}  + {n_I}{M_I})} {(\sum\limits_{i = 1}^{I - 1} {{n_i}{r_i}}  + {n_I}{p_I})}}} \right.
 \kern-\nulldelimiterspace} {(\sum\limits_{i = 1}^{I - 1} {{n_i}{r_i}}  + {n_I}{p_I})}},\\
{\tau _{I + 1}} = {{\sum\limits_{i = 1}^I {{n_i}{b_i}} } \mathord{\left/
 {\vphantom {{\sum\limits_{i = 1}^I {{n_i}{b_i}} } {\sum\limits_{i = 1}^I {{n_i}{r_i}} }}} \right.
 \kern-\nulldelimiterspace} {\sum\limits_{i = 1}^I {{n_i}{r_i}} }}
\end{array}\]

\subsubsection{Relationship Between EB and EC of Aggregate Flow}

Given the scenario with $I$ types of cloud services, we are able to derive the minimum buffer size $B$ to satisfy a delay constraint $D$ for all flows. Here is the derivation of $B$:

\begin{eqnarray}
&&B = {\sup _{s \ge 0}}\{ {\alpha ^*} - {e_D}({\alpha ^*})\} \nonumber \\
  &&= {\sup _{s \ge 0}}\{ \min (\sum\limits_{i = 1}^I {{n_i}({p_i}s + {M_i})} ,{\kern 1pt} {\kern 1pt}  \cdots {\kern 1pt} ,{\kern 1pt} {\kern 1pt} \sum\limits_{i = 1}^I {{n_i}({r_i}s + {b_i})} ) - {e_D}({\alpha ^*})\}\nonumber \\
  &&= {\sup _{s \ge 0}}\{ \min (\sum\limits_{i = 1}^I {{n_i}({p_i}s + {M_i})}  - {e_D}({\alpha ^*}),{\kern 1pt} {\kern 1pt}  \cdots {\kern 1pt} ,{\kern 1pt} \sum\limits_{i = 1}^I {{n_i}({r_i}s + {b_i}) - {e_D}({\alpha ^*})} )\}\nonumber \\
  &&= \max ({\sup _{0 \le s < {\Gamma _1}}}\{ \sum\limits_{i = 1}^I {{n_i}({p_i}s + {M_i})}  - {e_D}({\alpha ^*})\} ,{\kern 1pt}  \cdots {\kern 1pt} {\kern 1pt} ,{\kern 1pt} {\kern 1pt} {\kern 1pt} {\sup _{s \ge {\Gamma _I}}}\{ \sum\limits_{i = 1}^I {{n_i}({r_i}s + {b_i})}\nonumber \\
  &&- {e_D}({\alpha ^*})\} )\nonumber\\
  &&= \max (\sum\limits_{i = 1}^I {{n_i}{M_i}}  - {e_D}({\alpha ^*}),{\kern 1pt} {\kern 1pt} {\kern 1pt}  \cdots {\kern 1pt} {\kern 1pt} ,{\kern 1pt} {\kern 1pt} \sum\limits_{i = 1}^I {{n_i}({r_i}{\Gamma _I} + {b_i}) - {e_D}({\alpha ^*})} )
  \label{eq:2.8}
\end{eqnarray}

where, ${\alpha ^*} = \sum\limits_{i = 1}^I {{n_i}{\alpha _i}} $

\subsection{Admission Control of Aggregate Flow}

With the knowledge of characteristics of aggregate flow introduced previously, we now propose an admission control strategy.

\begin{equation}
{e_D}(\sum\limits_{i = 1}^I {{n_i}{\alpha _i}} ) \le C
\label{eq:2.9}
\end{equation}

This admission control strategy is essentially based on EB and we term this as EB-Based Admission Control, EBBAC. This means only cloud service requests that meet the admission control are accepted, otherwise they will be declined. For those accepted requests, the required buffer size are calculated as in Eq.(\ref{eq:2.8}) in Subsection 5.2.2. This ensures the delay constraint of all cloud services is less than $D$ with the same amount of accepted cloud services. A set of maximum values $({n_1},{n_2}, \cdots ,{n_I})$ that satisfies (\ref{eq:2.9}) represents the largest amount of cloud services a node can serve. This will be used to evaluate the admission control performance, where ${n_i} \in N$.

\section{Numerical Results}

In this section, extensive numerical experiments are conducted, from which we evaluate the characteristics of aggregate flows including EB, EC, the relationship between them and performance of EBBAC with various constraints. Parameters of $T$-$SPEC$, referred to \cite{le2001}, are summarized in Table \ref{tab1}.

\begin{table}[htbp]
 \centering
 \caption{\label{tab:test}Three Parameters for Cloud Services}

 \begin{tabular}{lccccr}
  \toprule  & ${p_i}$ & ${M_i}$ & ${r_i}$ & ${b_i}$ & ${\Gamma _i}$ \\\cline{2-6}
          $i$ & $Mb/s$ & $kb$ & $Mb/s$ & $kb$ & $ms$ \\
  \midrule 1 & 29 & 1 & 0.7 & 38 & 1.3 \\
           2 & 7 & 1 & 0.7 & 368 & 58.3 \\
           3 & 0.3 & 15 & 0.03 & 38 & 85.2 \\
  \bottomrule
 \end{tabular}
 \label{tab1}
\end{table}

\subsection{Characteristics of Aggregate Flow}

\begin{figure}[h]
  \centering
  \subfloat[\texttt{$D\;vs.\;EB\;for\;I=2$}]{\includegraphics[width=2in]{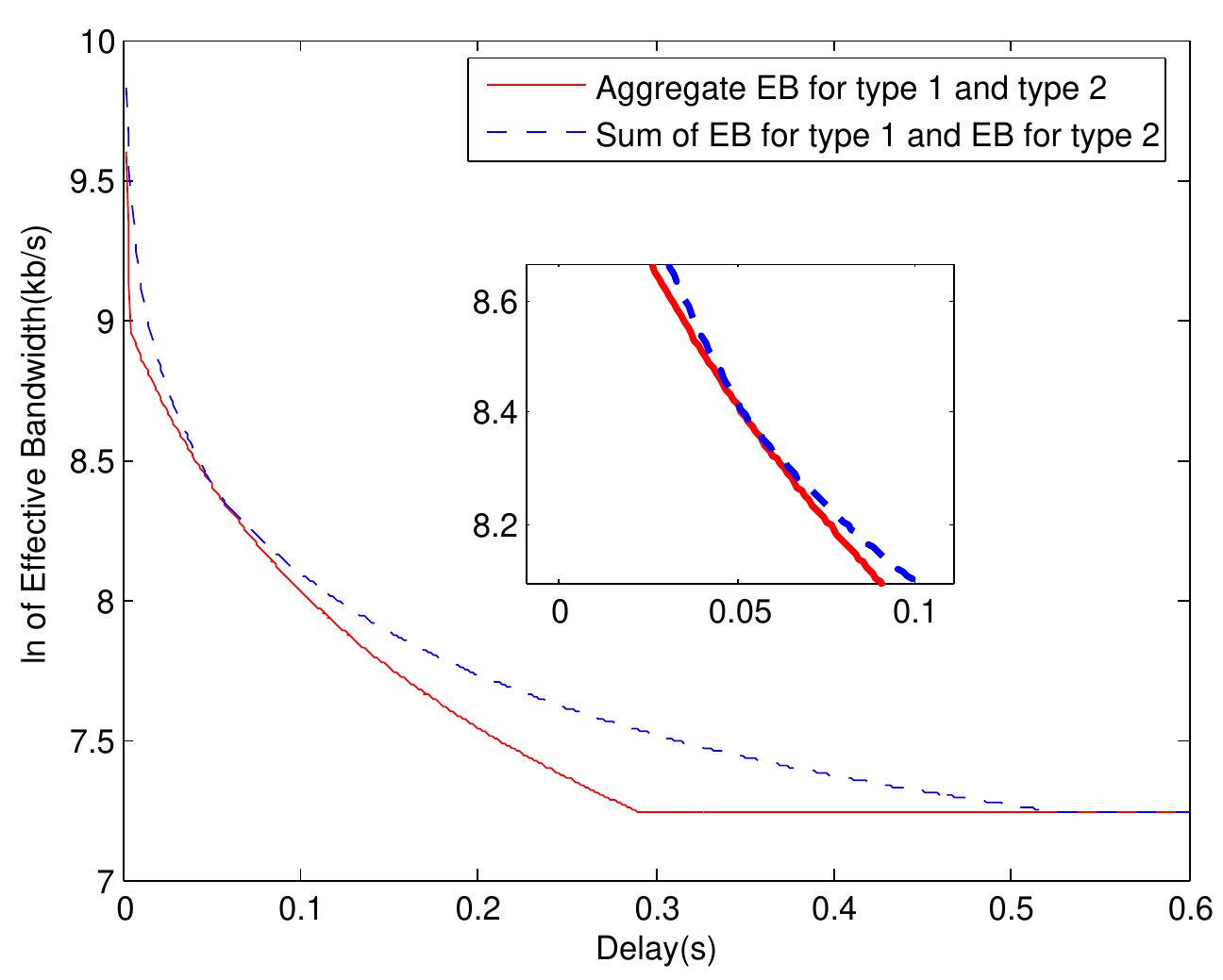}}
  \subfloat[\texttt{$D\;vs.\;EB\;for\;I=3$}]{\includegraphics[width=2in]{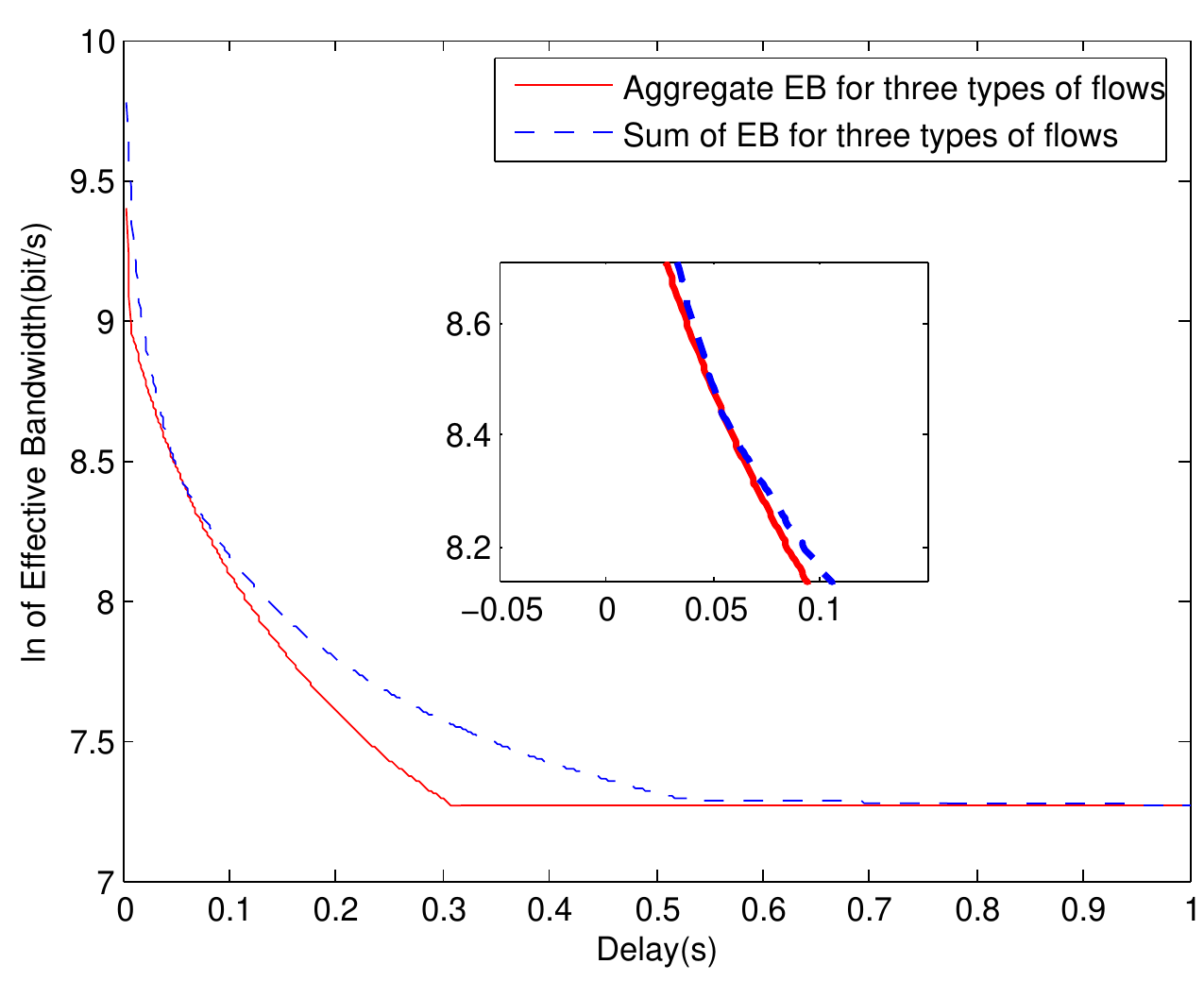}}\\
  \caption{Characteristics of EB for Aggregate Flow}
  \label{fig3}
\end{figure}

In the first experiment, we investigate the EB of aggregate flow and results are shown in Figure 3 (a) and Figure 3 (b). It can be seen that EB of the aggregate flow is less than or equal to the sum of EB from individual flows when there are two types ($I$=2) and three types ($I$=3) cloud services. This validates the accuracy of (\ref{eq:3.6}) derived in Section 3. In addition, the EB decreases when the delay constraint $D$ increases. This is because that a longer delay constraint needs a slower service rate, but the service rate cannot below the average rate of cloud flows. Therefore, when the delay constraint exceeds a specific value, the EB remains constant regardless of a further increase of the delay constraint. We also observe that the EB of the aggregate flow is equal to the sum of individual EBs, i.e., ${e_D}(\sum\nolimits_i {{\alpha _i}} ) = \sum\nolimits_i {{r_i}}  = \sum\nolimits_i {{e_D}({\alpha _i})}$, when $D \ge 0.53s$ in Figure 3(a) and $D \ge 1.27s$ in Figure 3(b).

\subsection{Relationship between EB and EC}

In the second experiment, we evaluate the relationship between EB and EC considering two scenarios: a single flow and aggregate flow.

\begin{figure}[h]
  \centering
  \subfloat[\texttt{$D\;vs.\;B\;for\;I=1$}]{\includegraphics[width=1.6in]{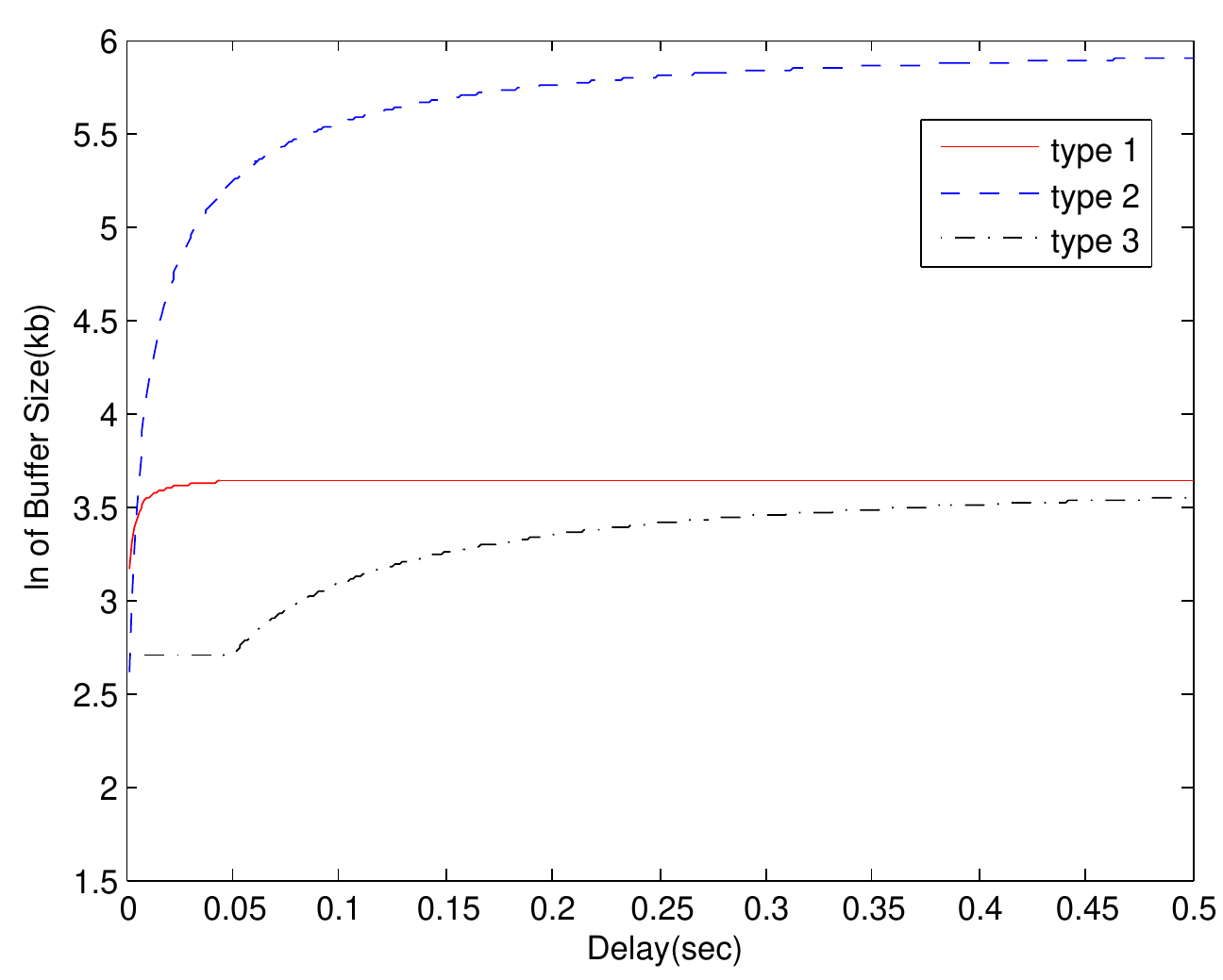}}
  \subfloat[\texttt{$D\;vs.\;B\;for\;I=2$}]{\includegraphics[width=1.6in]{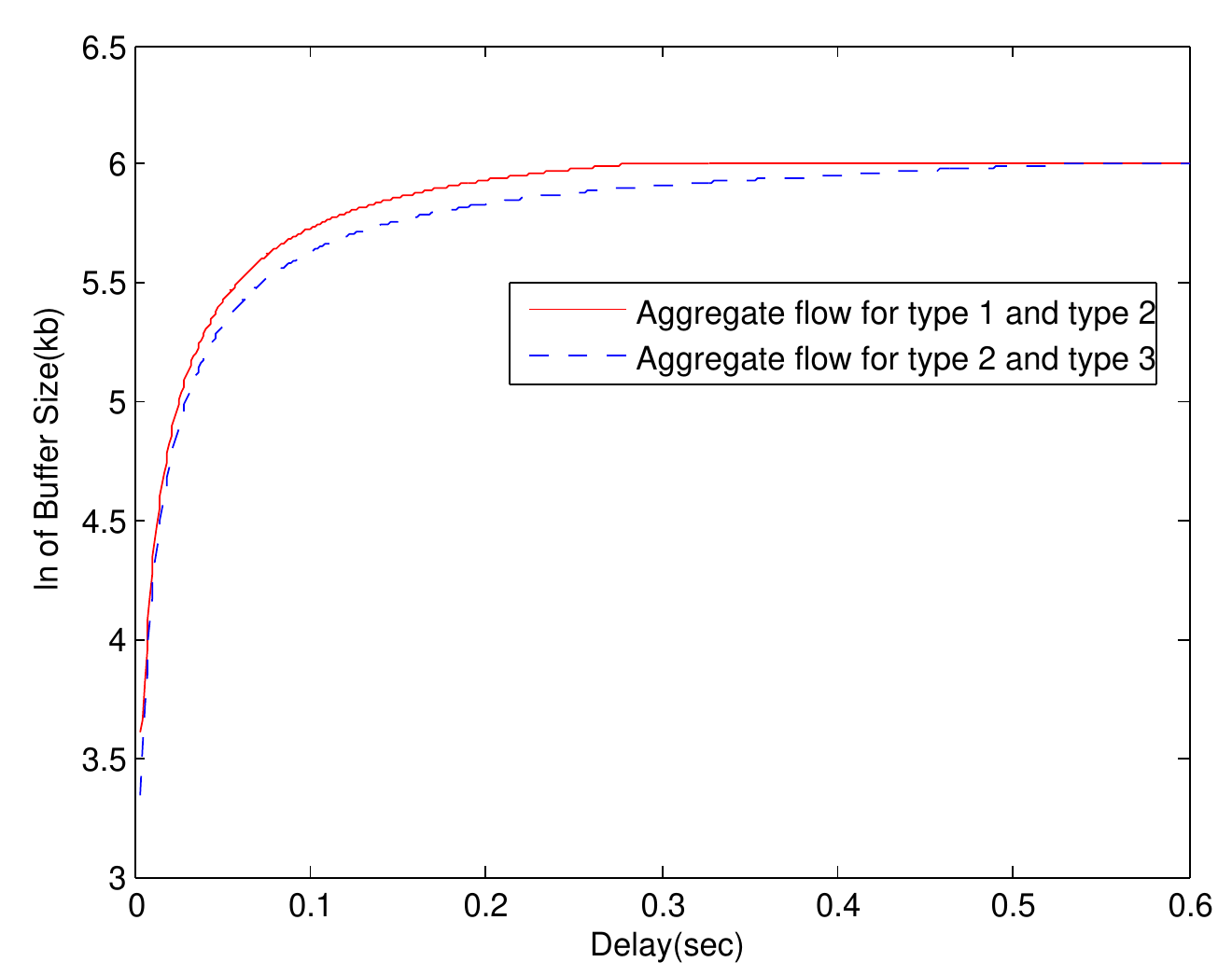}}
  \subfloat[\texttt{$D\;vs.\;B\;for\;I=3$}]{\includegraphics[width=1.6in]{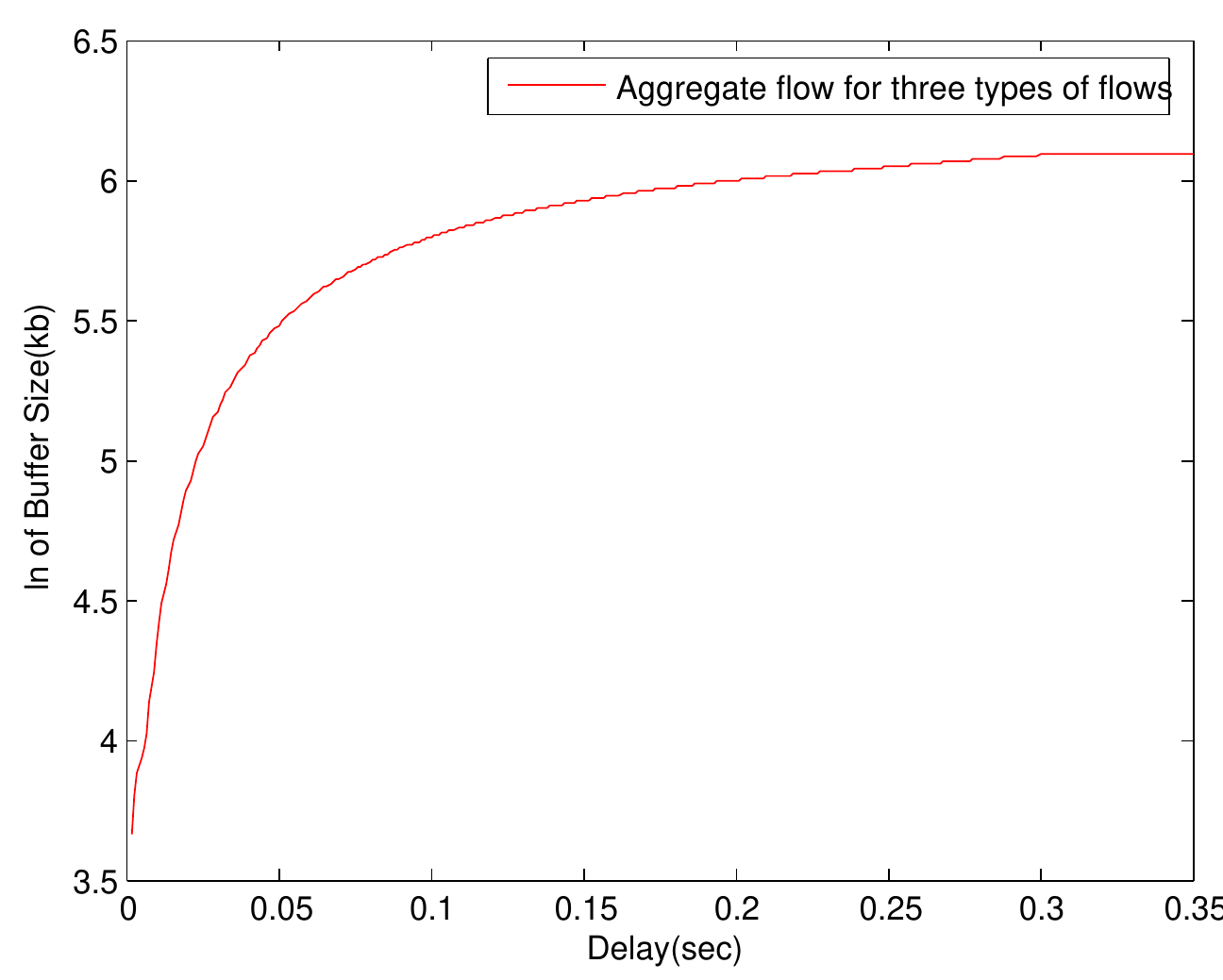}}\\
  \subfloat[\texttt{$D\;vs.\;B\;and$ EB $for\;I=1$}]{\includegraphics[width=1.63in]{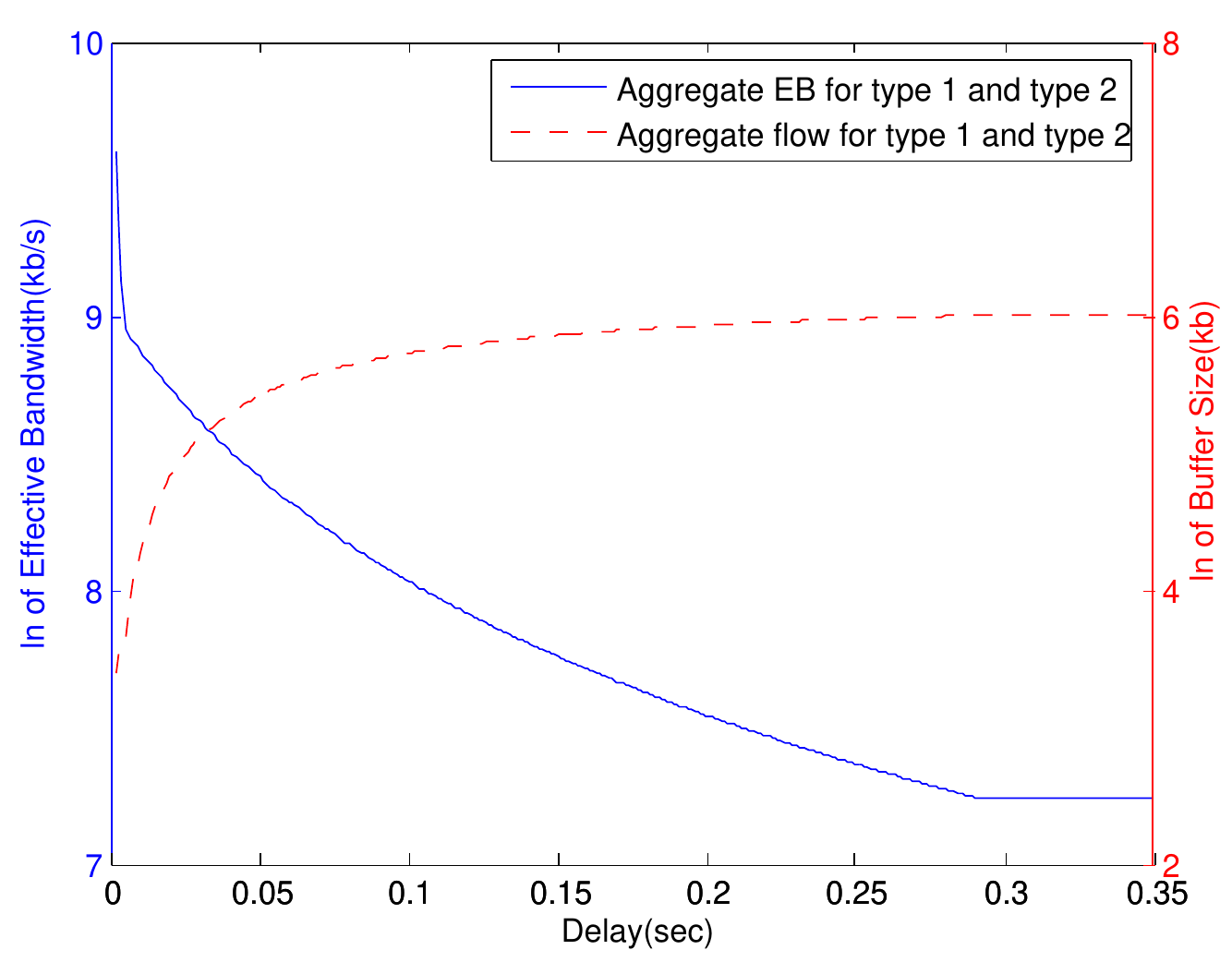}}
  \subfloat[\texttt{$D\;vs.\;B\;and$ EB $for\;I=2$}]{\includegraphics[width=1.63in]{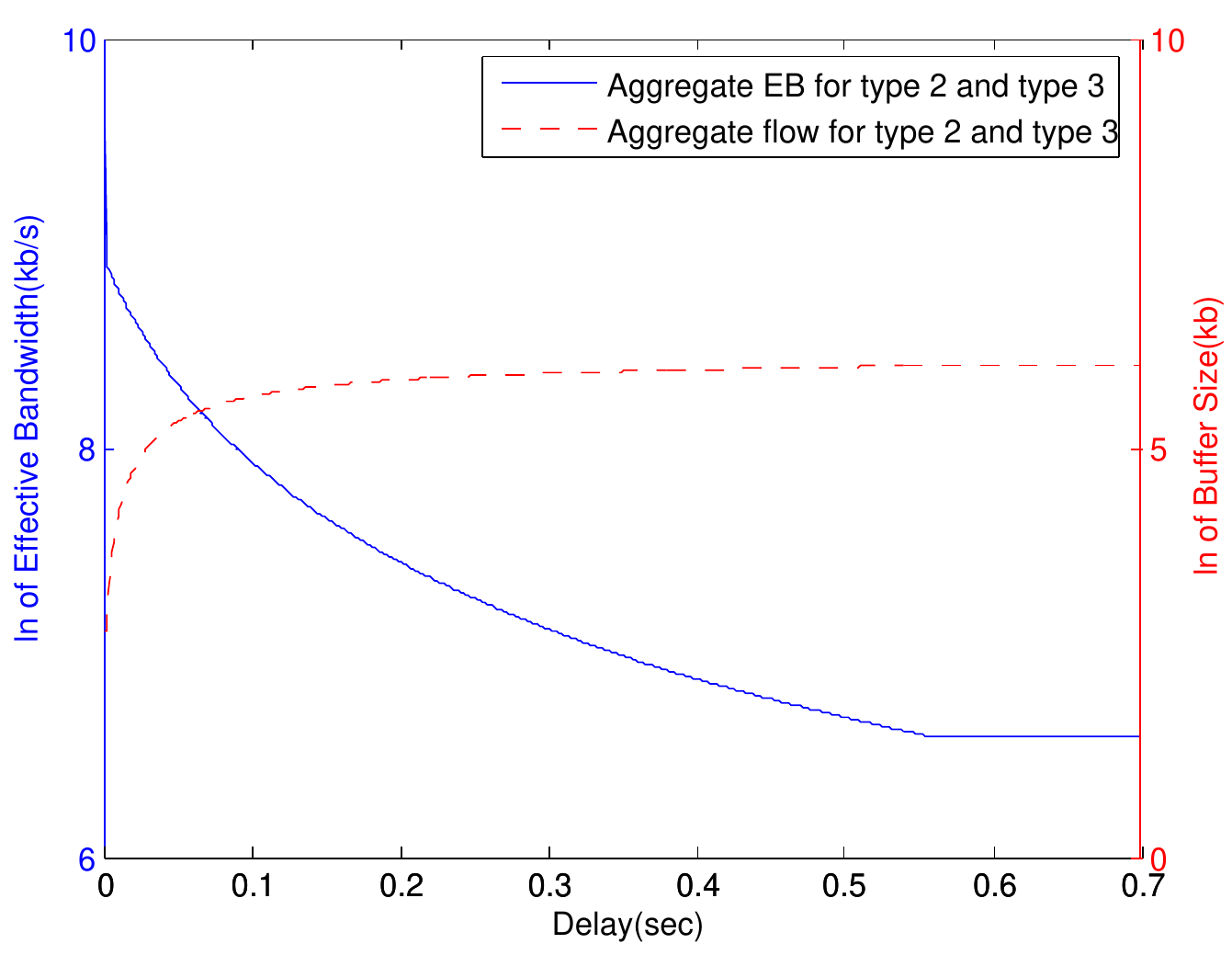}}
  \subfloat[\texttt{$D\;vs.\;B\;and$ EB $for\;I=3$}]{\includegraphics[width=1.63in]{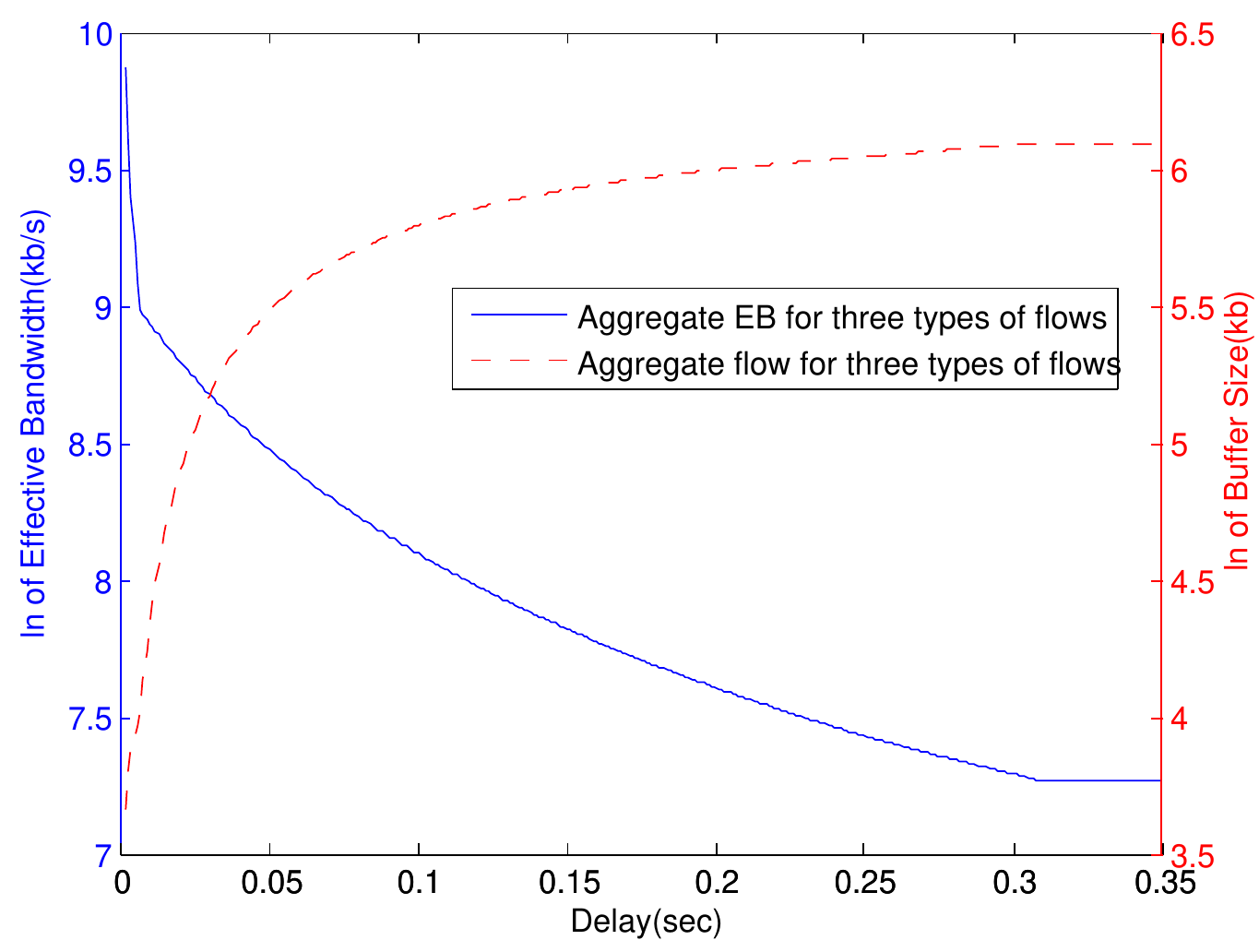}}
  \caption{Relationship between EB and EC}
  \label{fig4}
\end{figure}

The result for a single flow is plotted in Figure 5(a) and results for aggregate flows are described in Figure 5(b) and (c). All three figures demonstrate that the buffer size $B$ first increases along with $D$, then it remain constant. This matches the physical meaning, because the service rate offered for cloud flows reduces while $D$ rising, which leads to an increment of backlog. It further requires a larger buffer size $B$. However, the buffer size will eventually achieve a constant value when the service rate for flows equals the average rate of flows, because the backlog is no longer changed.

In the third experiment, we examine the relationship among the delay constraint $D$, effective bandwidth (EB) and the buffer size $B$ as shown in Figure 5(d), 5(e) and 5(f). Also, figures clearly show that the buffer size $B$ keeps unchanged when EB no longer increases regardless of the increment of $D$. This once again verifies the conclusion derived from the second experiment. That is, the buffer size will eventually achieve a constant value when the backlog is no longer changed because the service rate for flows equals the average rate of flows.

\subsection{Performance of Proposed AC Approach}

\begin{figure}[h]
  \centering
  \subfloat[\texttt{${n_1\;vs.\;n_2\;for\;I=2}$}]{\includegraphics[width=2in]{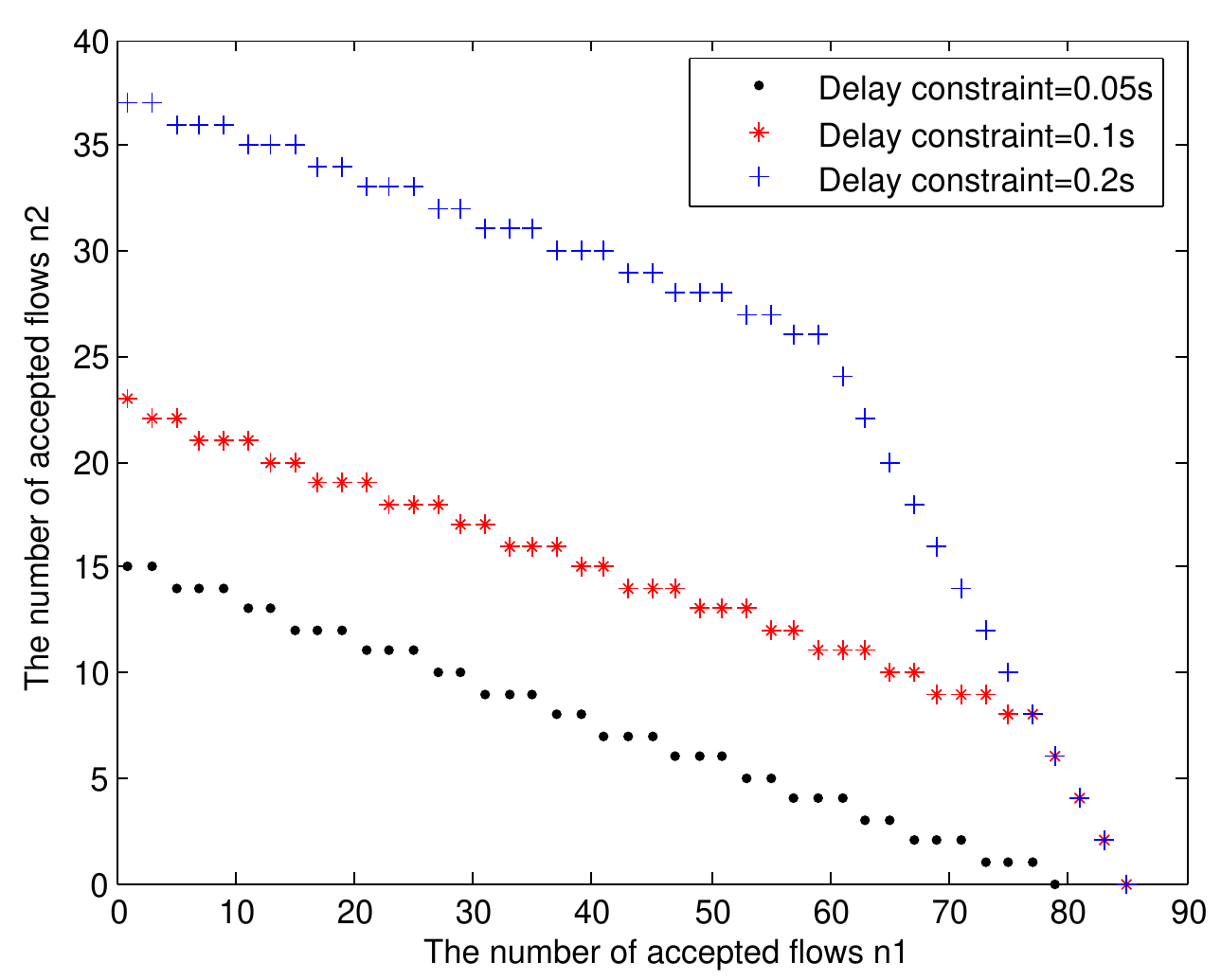}}
  \subfloat[\texttt{${n_2\;vs.\;n_3\;for\;I=2}$}]{\includegraphics[width=2in]{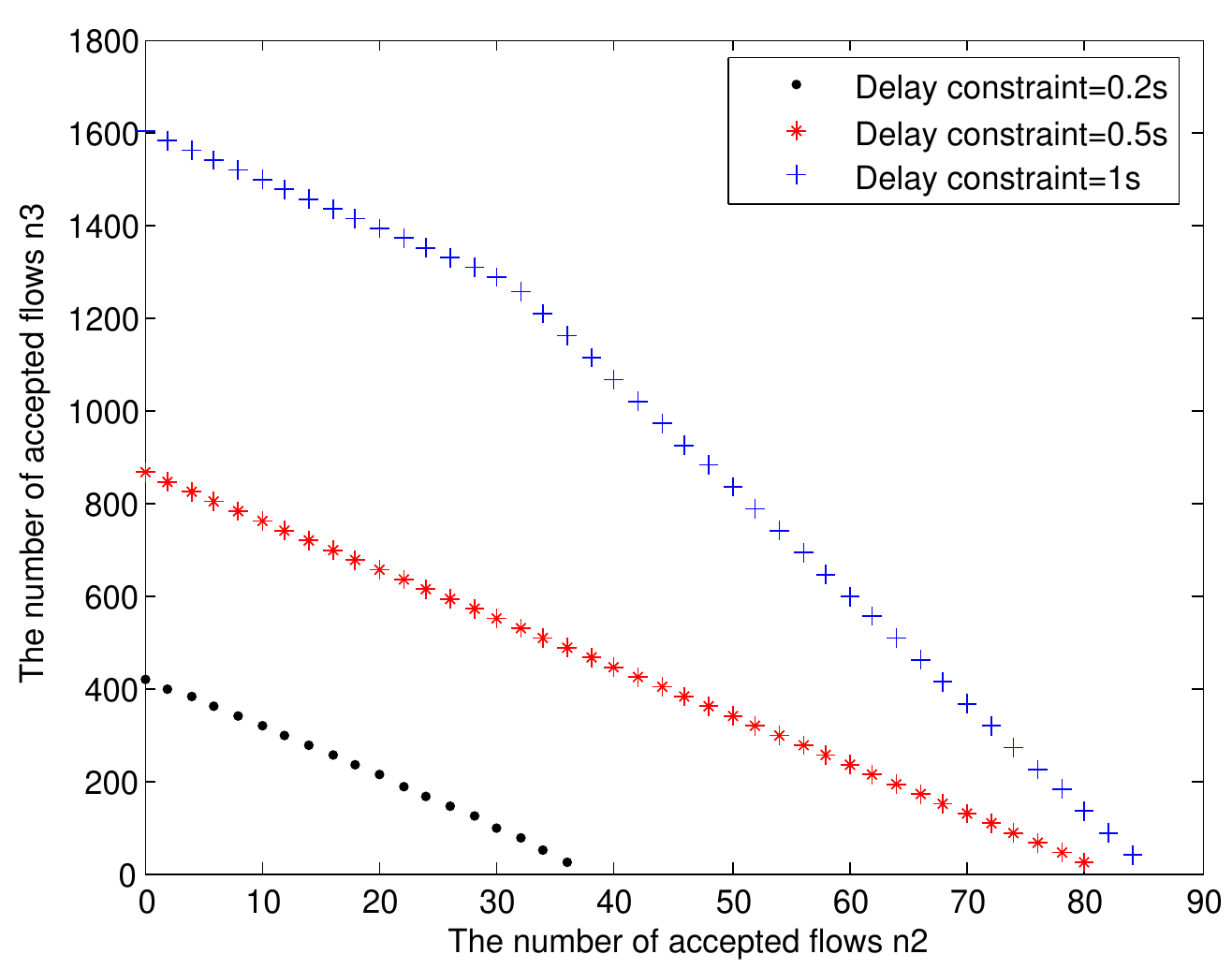}}\\
  \subfloat[\texttt{${n_1\;vs.\;n_3\;for\;I=2}$}]{\includegraphics[width=2in]{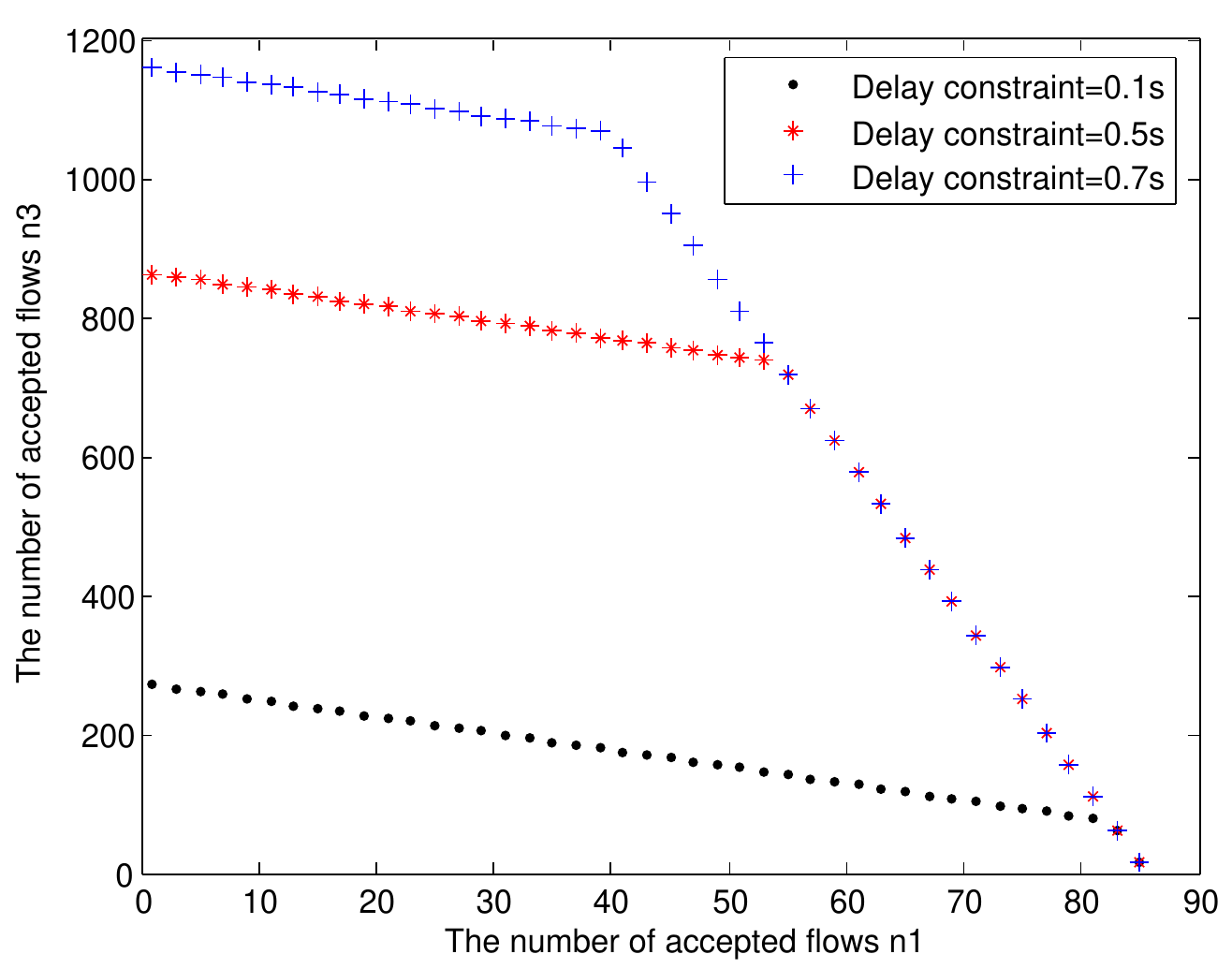}}
  \subfloat[\texttt{${n_1,\;n_2\;vs.\;n_3\;for\;I=3}$}]{\includegraphics[width=2in]{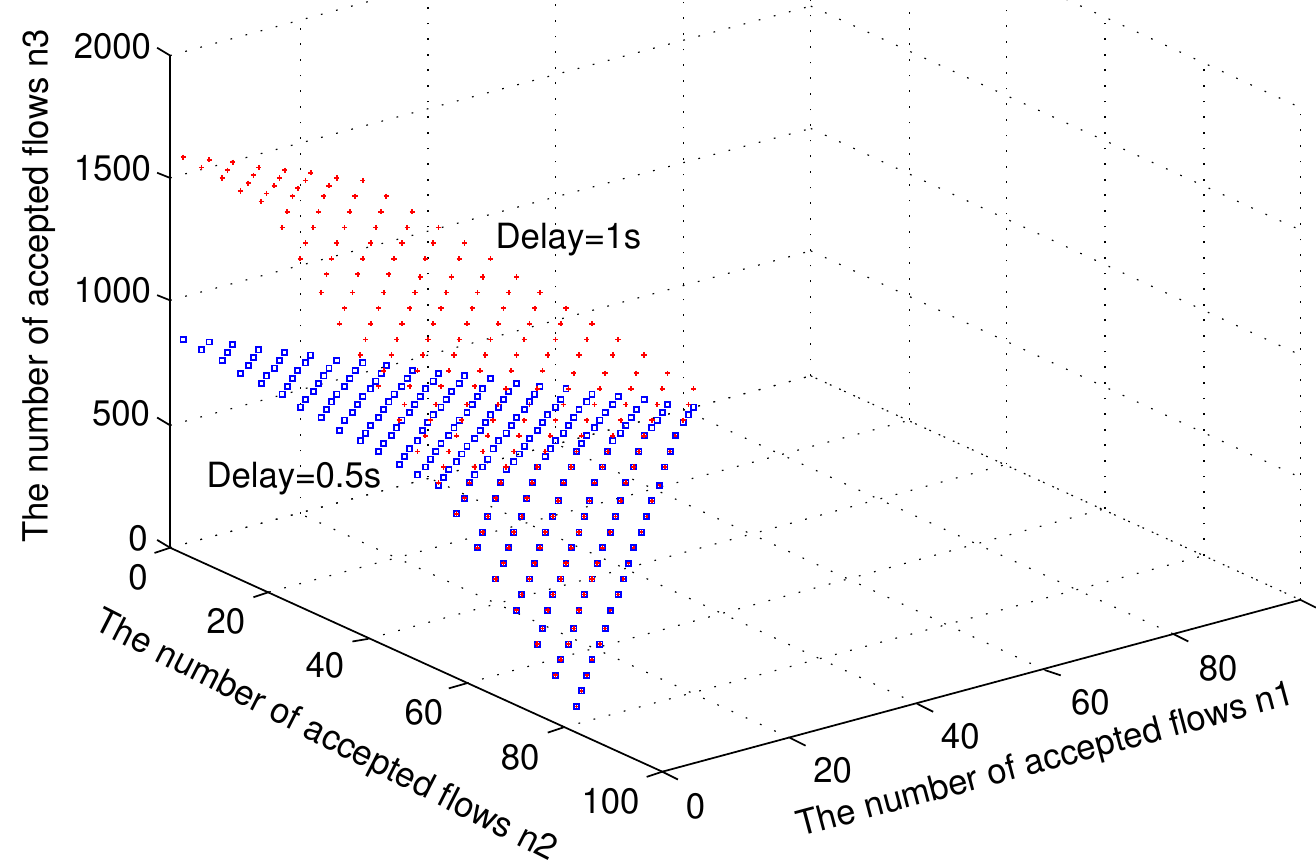}}
  \caption{The Performance of EBBAC}
  \label{fig5}
\end{figure}

In the last experiment, we study the performance of EB-Based Admission Control (EBBAC) and find out the maximum cloud services a system can support with various delay constraints $D$. Figures 5(a), (b) and (c) show the result with $I$=3 while Figure 5(d) with $I$=3, where $n_1,n_2,n_3$ are non-negative integers and represent the number of flows for three different type. From Figure 5, it can be observed the node can accept a larger amount of clouds flows with a bigger delay constraint. As mentioned previously, cloud flows needs a slower service rate when the delay constraint goes up. Therefore, it is able to serve more cloud flows with a fixed output rate $C$.

Furthermore, Eq. (19) indicates that the effective bandwidth varies with the maximum amount of accepted cloud flows. From Figure 5(a), we find that $n_1$ (when $n_2$=0) of the first type is greater than that $n_2$ (when $n_1$=0) of the second type. Note that this may not always hold due to the selection of the delay constraint $D$. In in our experiments, we arbitrarily choose a set of values for $D$ and generate several more figures in Figure 5(b), (c) and (d), which reveal the similar characteristics with Figure 5(a).

\section{Conclusions}

In this paper we proposed a model of Cloud-oriented call admission control to guarantee QoS provisioning for Cloud services. In order to address the challenges brought in by large number of parallel service requests in Clouds, we developed an admission control scheme for aggregate flow. We employed network calculus to determine effective bandwidth for aggregate flow for making admission decisions. In order to improve network resource allocation as well as provide QoS guarantee for Cloud services, we also examined relationship between effective bandwidth and equivalent capacity. We also reported extensive experimental results to show features of effective bandwidth, verify the relation between effective bandwidth and equivalent capacity, and evaluate performance of the proposed admission control scheme under various delay upper bound requirements.

\begin{flushleft}
  \Large
  \textbf{Acknowledgments}
\end{flushleft}

This work is supported by the program New Century Excellent Talents in University and the following grants: National Science Foundation of China (Grant No. 61272400, 61309031), Chongqing Innovative Team Fund for College Development Project (Grant No. KJTD201310), Natural Science Foundation of Chongqing (Grant No. cstc2013jcyjA40026), The Research Project of Chongqing Education Committee (Grant No. KJ130523), and Chongqing University of Posts and Telecommunications Research Fund for Young Scholars (Grant No. A2012-79).

\bibliographystyle{unsrt}
\bibliography{citation}

\end{document}